\documentclass[conference]{IEEEtran}
\IEEEoverridecommandlockouts

\usepackage{cite}
\usepackage{amsmath,amssymb,amsfonts}
\usepackage{algorithmic}
\usepackage{graphicx}
\usepackage{textcomp}
\usepackage{xcolor}

\usepackage[utf8]{inputenc}
\usepackage[T1]{fontenc}
\usepackage{hyperref}
\usepackage{graphicx}
\usepackage{subcaption}
\usepackage{xcolor}
\usepackage{fancyvrb}
\usepackage{listings}
\usepackage{verbatim} 
\usepackage{url}
\lstdefinestyle{common}{
    xleftmargin=.5em,
    xrightmargin=.5em,
    frame=single,framesep=.5em,framerule=0pt,
    fancyvrb=true,
    basicstyle=\ttfamily\scriptsize,
    keywordstyle=\color{cyan!50!blue!75!black}\bfseries,
    commentstyle=\color{red!50!black}\itshape,
    stringstyle=\ttfamily\color{green!50!black},
    numbers=left,
    numberstyle=\tiny,
    showspaces=false,
    showstringspaces=false,
    fontadjust=true,
    keepspaces=true,
    flexiblecolumns=true,
    emphstyle=\color{red},
    breaklines=true,
}
\lstdefinestyle{log}{
    style=common,
    backgroundcolor=\color{blue!5},
    aboveskip=5pt,
    belowskip=5pt,
    language=bash,
    morestring=[d]{EOF},
    otherkeywords = {wormpi-mr, in, out, wormhole-in, provider, wormhole-out},
    fancyvrb=true,
    morekeywords = [2]{DP3T_ScanResponse, INFO},
    keywordstyle = [2]{\color{green!50!black}},
    morekeywords = [3]{fd68},
    keywordstyle = [3]{\color{red!50!black}},
    breaklines=true,
}
\lstdefinestyle{console}{
    style=common,
    backgroundcolor=\color{gray!10},
    aboveskip=5pt,
    belowskip=5pt,
}

\begin{document}

\title{Mind the GAP: Security \& Privacy Risks of Contact Tracing Apps}
\author{
\IEEEauthorblockN{
    Lars Baumgärtner\IEEEauthorrefmark{1},
    Alexandra Dmitrienko\IEEEauthorrefmark{3}, 
    Bernd Freisleben\IEEEauthorrefmark{2},
    Alexander Gruler\IEEEauthorrefmark{1},
    Jonas Höchst\IEEEauthorrefmark{1}\IEEEauthorrefmark{2},
}
\IEEEauthorblockN{
    Joshua Kühlberg\IEEEauthorrefmark{1},
    Mira Mezini\IEEEauthorrefmark{1},
    Richard Mitev\IEEEauthorrefmark{1},
    Markus Miettinen\IEEEauthorrefmark{1},
    Anel Muhamedagic\IEEEauthorrefmark{1},
    Thien Duc Nguyen\IEEEauthorrefmark{1},
}
\IEEEauthorblockN{
    Alvar Penning\IEEEauthorrefmark{2},
    Dermot Pustelnik\IEEEauthorrefmark{1},
    Filipp Roos\IEEEauthorrefmark{3},
    Ahmad-Reza Sadeghi\IEEEauthorrefmark{1},
    Michael Schwarz\IEEEauthorrefmark{2},
    Christian Uhl\IEEEauthorrefmark{2}
}
\IEEEauthorblockA{\IEEEauthorrefmark{1}
    Technische Universität Darmstadt, Germany\\
    E-mail: \{baumgaertner, mezini, markus.miettinen, ducthien.nguyen, ahmad.sadeghi\}@cs.tu-darmstadt.de \\
}
\IEEEauthorblockA{\IEEEauthorrefmark{2}
    Philipps-Universität Marburg, Germany\\
    E-mail: \{hoechst, freisleb, penning, schwarzx, uhlc\}@informatik.uni-marburg.de \\
}
\IEEEauthorblockA{\IEEEauthorrefmark{3}
    Julius-Maximilians-Universität Würzburg, Germany\\
    E-mail: \{alexandra.dmitrienko, filipp.roos\}@uni-wuerzburg.de \\
}
}

\maketitle
\begin{abstract}
Google and Apple have jointly provided an API for exposure notification in order to implement decentralized contract tracing apps using Bluetooth Low Energy, the so-called "Google/Apple Proposal", which we abbreviate by "GAP". 
We demonstrate that in real-world scenarios the current GAP design is vulnerable to (i) profiling and possibly de-anonymizing infected persons, and (ii) relay-based wormhole attacks that basically can generate fake contacts with the potential of affecting the accuracy of an app-based contact tracing system. 
For both types of attack, we have built tools that can easily be used on mobile phones or Raspberry Pis (e.g., Bluetooth sniffers). 
The goal of our work is to perform a reality check towards possibly providing empirical real-world evidence for these two privacy and security risks.
We hope that our findings provide valuable input for developing secure and privacy-preserving digital contact tracing systems. 
\end{abstract}


\section{Introduction}
Caused by coronavirus SARS-CoV-2, the COVID-19 disease spreads particularly through direct contact between people. 
Health authorities face the challenge of identifying and isolating infection chains to prevent the pandemic from spreading further. 
This task usually involves manual effort and relies on contact information that is voluntarily provided by infected people. 
Hence, infection chains have to be reconstructed by the health authorities with an enormous amount of work for each individual case, but nevertheless they are not always accurate or complete.

Using digital contact tracing apps on mobile devices can help to reduce the manual effort and increase the tracing accuracy. This has already been demonstrated in several countries. Even if - for understandable reasons - there is still a lack of empirical evidence about the effectiveness of contract tracing apps in fighting the pandemic \cite{BRAITHWAITE2020} compared to other measures, such as massive testing and manual tracing, contact tracing apps became urgent in the rest of the world that was hit by the pandemic later. 

The recent "arms race" on providing contact tracing apps has created various mobile contact tracing approaches and corresponding apps,  published in GitHub repositories, or deployed in countries such as China, South Korea, Singapore, Taiwan, Austria, Australia, France, Italy, Switzerland, UK, and Germany. 
For an overview and a comparison of contact tracing apps, we refer to Miettinen et al.~\cite{TraceCORONAcomparison}.

Basically, contact tracing apps differ in (a) the technology used to measure proximity, and (b) the approach of where and how contacts are stored and processed. For example, some approaches are based on tracking the GPS location of participating users \cite{PrivateKitMIT,AarogyaSetuIndia}, while other proposals rely on recording proximity identifiers exchanged by Bluetooth Low Energy (BLE) technology. Location information, e.g., GPS in an urban environment, is often inaccurate \cite{10.1371/journal.pone.0219890} and mostly useless in indoor situations where  BLE contact tracing is supposed to show its value: examples are department stores, restaurants, trade fairs, conference venues, concert halls, indoor sports events, airports, and airplanes. However, Leith and Farrell \cite{DBLP:journals/corr/abs-2006-06822,DBLP:journals/corr/abs-2006-08543,10.1371/journal.pone.0239943} indicate that BLE received signal strength can vary substantially depending on the relative orientation of handsets, on absorption by the human body, reflection/absorption of radio signals in buildings, busses, and trams.
In particular, the authors demonstrate that BLE signal strength in a bus can be higher between phones that are far apart than phones close together, making reliable proximity detection based on signal strength difficult.

Furthermore, there are centralized and decentralized contact tracing approaches. Centralized approaches, such as the TraceTogether app used in Singapore~\cite{TraceTogether} or the CovidSafe app used in Australia~\cite{CovidSafe}, as well as the PEPP-PT model originally planned to be deployed in Germany~\cite{PeppPT}, and the StopCovid app in France~\cite{StopCovidFrance}, are based on the principle that a centralized back-end server system assigns a unique identifier (a pseudonym) to each user’s mobile app, from which frequently-changing pseudonymous proximity identifiers are derived and broadcast over BLE to other nearby devices. The weakness of this approach is that the back-end server system can associate all proximity identifiers of all users with the unique pseudonym of each user. This in turn allows operators of the back-end server system to perform comprehensive monitoring of all users of the system.

Due to these severe weaknesses, contact tracing apps should be based on decentralized identification of contacts. In the decentralized approach, the back-end server system does not have information about proximity identifiers of users and therefore cannot associate them with individual users. Indeed, due to the fundamental problems associated with the centralized approach and the huge resistance of the scientific community\footnote{Joint Statement on Contact Tracing: Date 19th April 2020,  \url{https://drive.google.com/file/d/1OQg2dxPu-x-RZzETlpV3lFa259Nrpk1J/view}}, the Federal Government of Germany decided to abandon the centralized approach of the PEPP-PT consortium that it initially highly supported, and decided to switch to a decentralized approach.

Apart from proposals made by academia, IT enterprises such as Microsoft, Apple, and Google work on contact tracing apps. In an unprecedented joint effort, Google and Apple have provided an Application Programming Interface (API) for exposure notification at the mobile operating system level in order to implement decentralized contract tracing apps using BLE. We call this the \textit{Google/Apple Proposal}~\cite{ag2020exposure:crypto}, abbreviated by GAP in the rest of the paper.

Several researchers have pointed out the possibility of profiling infected persons in the GAP approach \cite{chan2020pact,gvili2020security}, as well as the possibility to perform relay attacks \cite{eu2020mobile,chan2020pact, vaudenay2020analysis,vaudenay2020centralized,pietrzak2020delayed}. 
The goal of our work is to perform a reality check towards possibly providing empirical real-world evidence for these two privacy and security risks 
discussed in the literature.

We selected the GAP approach for the following reasons. First, GAP will be broadly adopted, since several European contact tracing apps, such as the Swiss \textit{SwissCOVID}, the Italian \textit{Immuni}, and the German \textit{Corona-Warn-App}, are based on the GAP API. 
Second, the GAP API is already opt-in for iOS and Android devices, hence, it will potentially stay with us for a long time. 

We demonstrate that in real-world scenarios the current GAP design is vulnerable to (i) profiling and possibly de-anonymizing infected persons, and (ii) relay-based wormhole attacks that can generate fake contacts with the potential of affecting the accuracy of an app-based contact tracing system. For both types of attack, we have built tools that can be easily used on mobile phones or Raspberry Pis (e.g., Bluetooth sniffers).  
We hope that our findings provide valuable input in the process of testing and certifying contact tracing apps, 
ultimately guiding improvements for secure and privacy-preserving design and implementation of digital contact tracing systems. 

The paper is organized as follows. Section \ref{gap} briefly explains the GAP approach. In Section \ref{profiling}, we present our (profiling) attack on the privacy of the GAP scheme. Section \ref{wormhole} describes our (relay-based wormhole) attack on the security of GAP. Section \ref{conclusion} concludes the paper and outlines areas for future work.

\section{GAP Overview}
\label{gap}

\begin{figure}
    \centering
    \includegraphics[width=1.0\columnwidth]{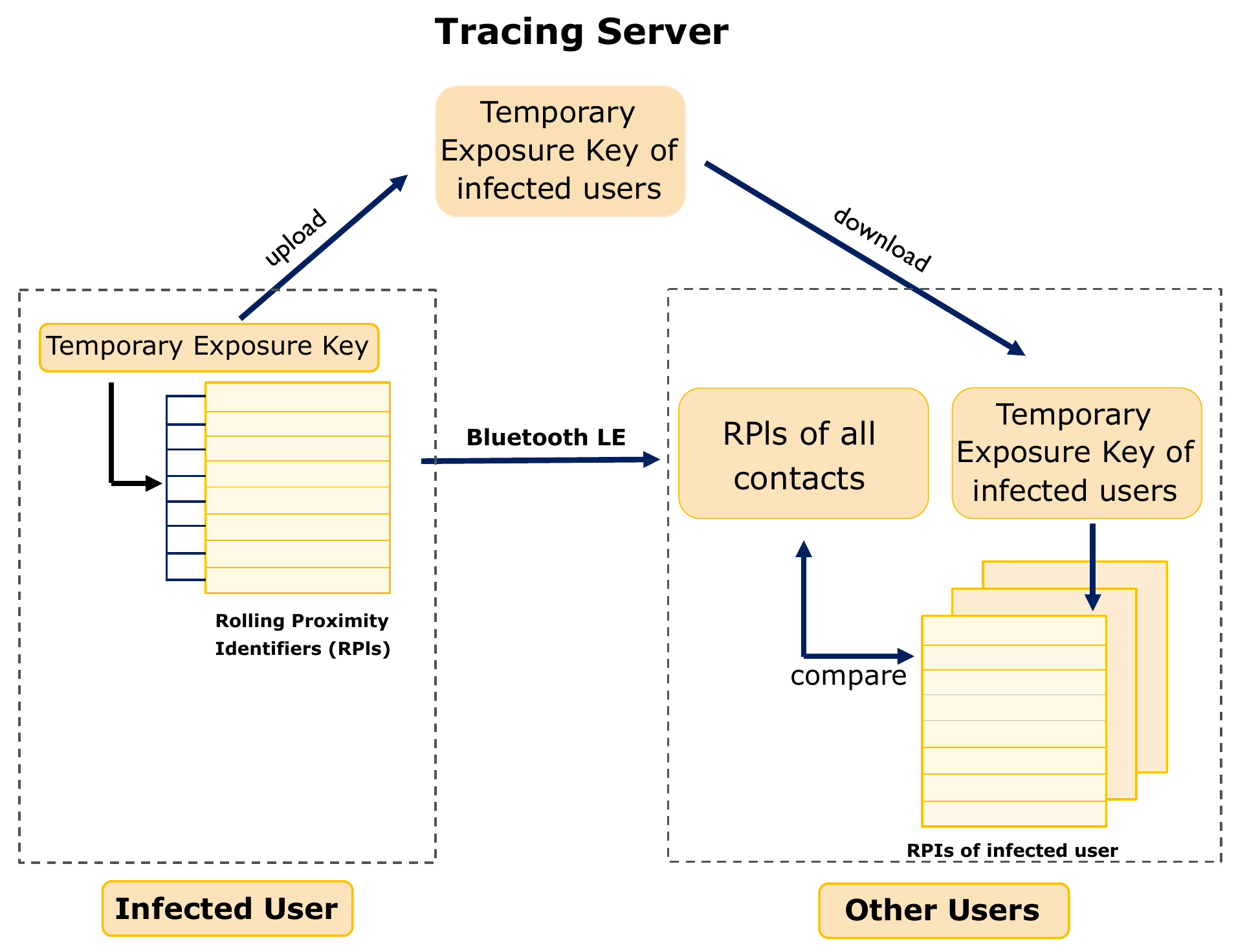}
    \caption{Overview of the GAP contact tracing approach}
    \label{fig:GAP-overview}
\end{figure}

The GAP contact tracing approach \cite{ag2020exposure:crypto} is based on frequently-changing random pseudonyms, so-called \emph{Rolling Proximity Identifiers (RPI)}. An overview of the approach is shown in Fig.~\ref{fig:GAP-overview}. Each app generates these RPIs from a \emph{Temporary Exposure Key (TEK)} (formerly known as \emph{Daily Tracing Key (DTK)} in version 1.0 of the GAP specification) and beacons them into their surroundings using BLE. Apps on other devices in close proximity can observe these RPIs and store them locally as a record of contact with the device beaconing the RPI. This dataset also includes additional metadata like the received signal strength. 

Should a user be tested positive for SARS-CoV-2, a user can decide to upload TEKs of the last $x$ days using an app to a central server (currently $x=14$). The server accumulates the received TEKs of infected persons and offers them to be downloaded by other users' apps.
Apps in devices of participating users regularly check the server for updates and download any new TEKs. Each app then uses the downloaded TEKs to calculate the corresponding RPI pseudonyms used by the infected persons' apps in the recent past. The operating system / corresponding system service then compares these infected persons' RPIs to the RPIs stored locally on the device. If matching RPIs are found, the metadata, e.g., signal strength or duration of the encounters, related to these matching RPIs are used to calculate a risk score that is used to determine whether a warning should be displayed to the user or not.

\section{Mind the Privacy GAP}
\label{profiling}

In this section, we present our real-world attack on the privacy of the GAP scheme based on its specification~\cite{ag2020exposure:crypto}. 
The possibility of profiling infected persons in the GAP has already been pointed out by other researchers \cite{chan2020pact,gvili2020security}.
The main problem is that the GAP requires infected individuals to publish all BLE proximity identifiers they have used during the days they may have been infectious to all devices participating in the system. Thus, this information essentially becomes public, and we argue that it is possible for all app users and other entities involved in the system to potentially track the movements of (and possibly de-anonymize~\cite{ji2015USENIXSecGraph, Radaelli2018SurveillanceNetworkedAge}) infected persons during these days. 

\subsection{Privacy Attack: Profiling Infected Persons}
\label{sec:profiling-attack}

\subsubsection{Goal and System Setup} 
The goal of our experiment is to show that it is practically possible to profile the movement and activities of infected users after they upload their TEKs.
Based on TEKs, other participating apps can derive the corresponding RPIs that the infected user's app has beaconed out in the recent past. 
Note that since all TEKs uploaded by infected persons can be downloaded by anyone, the RPIs are essentially public information.

\begin{figure}[ht]
    \centering
    \includegraphics[width=.85\columnwidth]{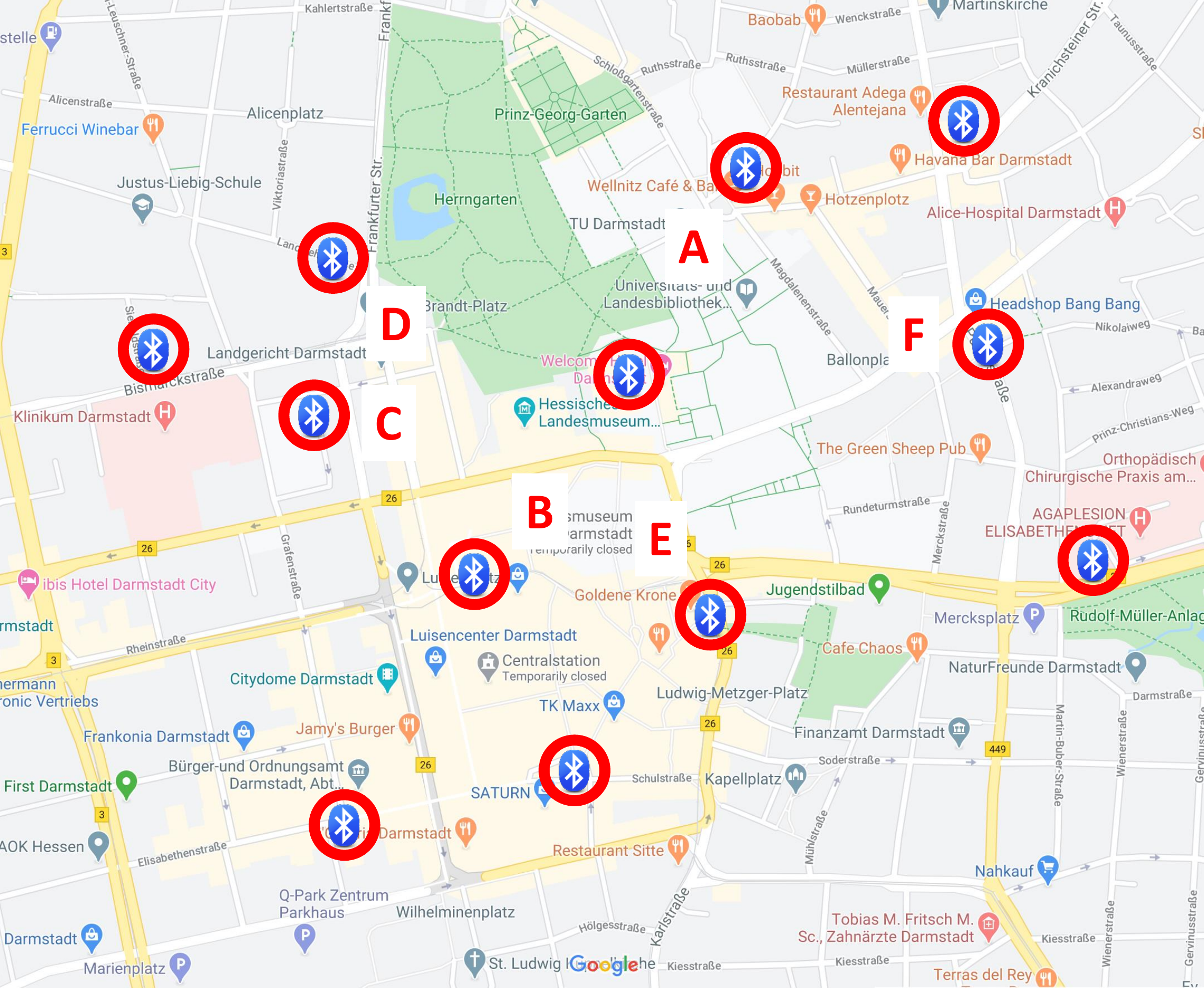}
    \caption{An example of observation points}
    \label{fig:profiling-attack-deploy}
\end{figure}

To conduct the attack, we deployed BLE sniffers to capture RPIs at six selected sensitive places downtown the city of Darmstadt, Germany, as listed in Table \ref{tab:list-location}. The locations of these places in Darmstadt are shown in Fig.~\ref{fig:profiling-attack-deploy}. We used commodity smartphones as the sniffers that can capture BLE signals at a distance of up to 6 meters. However, with a special Bluetooth antenna it was possible to capture signals at a much higher distance. The BLE sniffers would capture RPIs of any users moving through or spending time at the places mentioned above. In our experiment, two tracing app users simulate two particular paths. 

\begin{table}[ht]
\centering
\caption{List of locations with deployed BLE sniffers}
\label{tab:list-location}
\renewcommand{\arraystretch}{1.2}
\begin{tabular}{ll}
Location & Description    \\\hline
A        & A residential area        \\
B        & City hall      \\
C        & Police station \\
D        & Clinic and pharmacy     \\
E        & Outside a pub            \\
F        & Outside a head shop and a sports gambling bookmaker
\end{tabular}
\end{table}

Since the official GAP API can currently only be used by governmental health institutions \cite{AppleExposureAddendum}, we implemented a GAP tracing app simulator, following the RPI generation procedure laid out in the GAP cryptography API specification~\cite{ag2020exposure:crypto}.

\subsubsection{Experimental Results}
\label{sec:profiling-experiments}
\begin{figure}[ht]
	\centering
			\includegraphics[width=1.00\columnwidth]{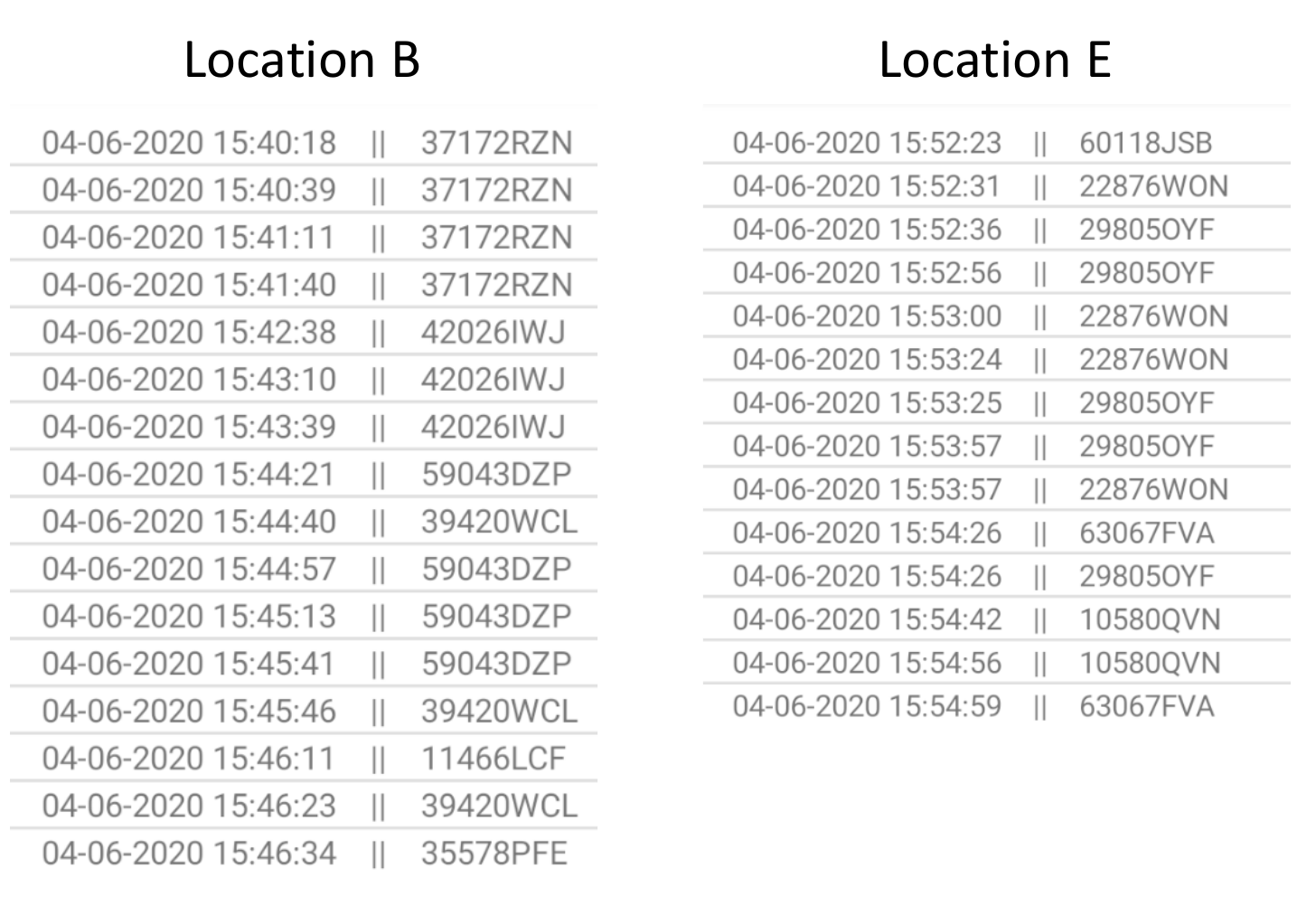}
        \caption{RPI measurements captured at location B and E}
        \label{fig:profiling-attack-capture}
\end{figure}

\begin{figure}[ht]
    \centering
    \includegraphics[width=\columnwidth]{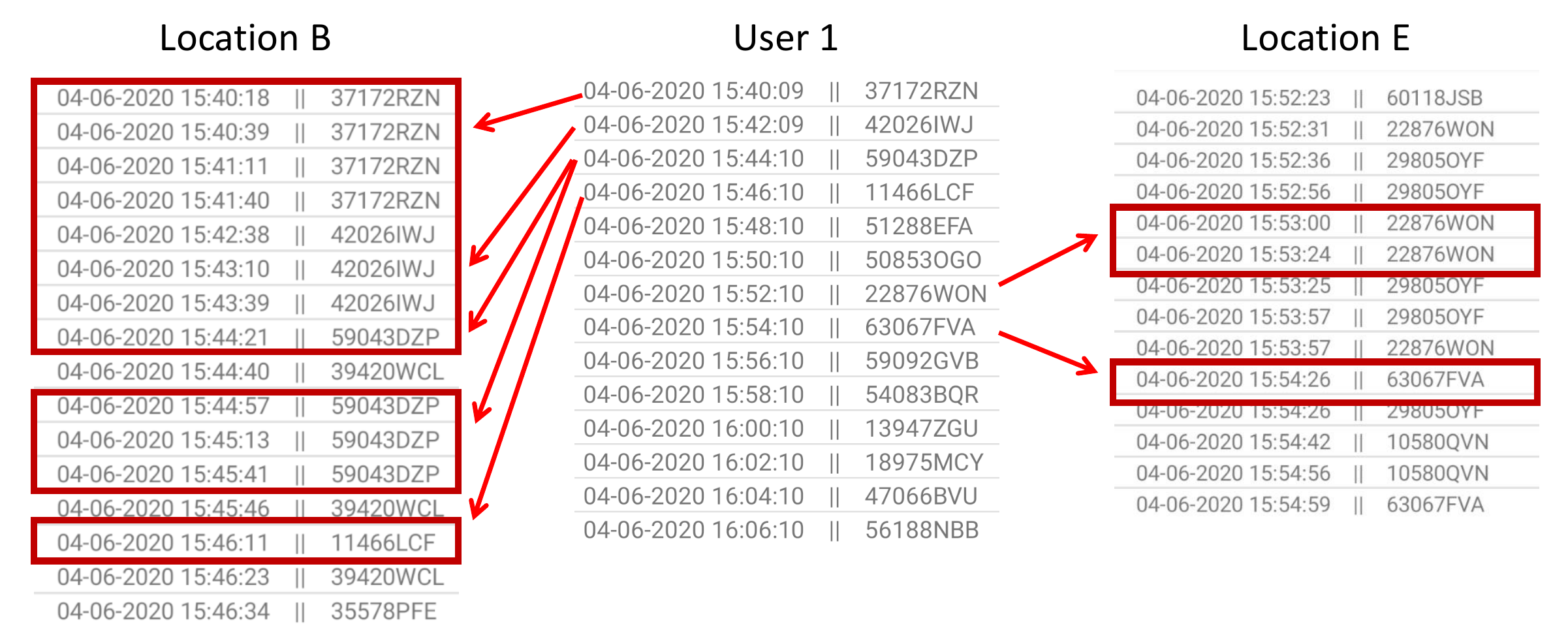}	
	\caption{Profiling User 1 movements}
    \label{fig:profiling-attack-link}
	\label{fig:profiling-attack-measurement}
\end{figure}

A sample of our results of RPI measurements captured at different observation points (marked from A to F) is shown in Fig.~\ref{fig:profiling-attack-capture}. The captured data looks entirely random, and it is not obvious which RPIs could be associated with individual users. However, when we simulate the case that any one of the users is tested positive for SARS-CoV-2 and these users upload their TEKs that were used to derive the corresponding RPIs, a completely different picture emerges, as shown in Fig.~\ref{fig:profiling-attack-link}.  It is evident that by matching the RPIs of User 1 with the RPIs captured in different locations, e.g., location $B$ and location $E$, we know exactly which locations User 1 has visited and when User 1 arrived and left each location.

\begin{figure}[ht]
    \centering
    \includegraphics[width=.85\columnwidth]{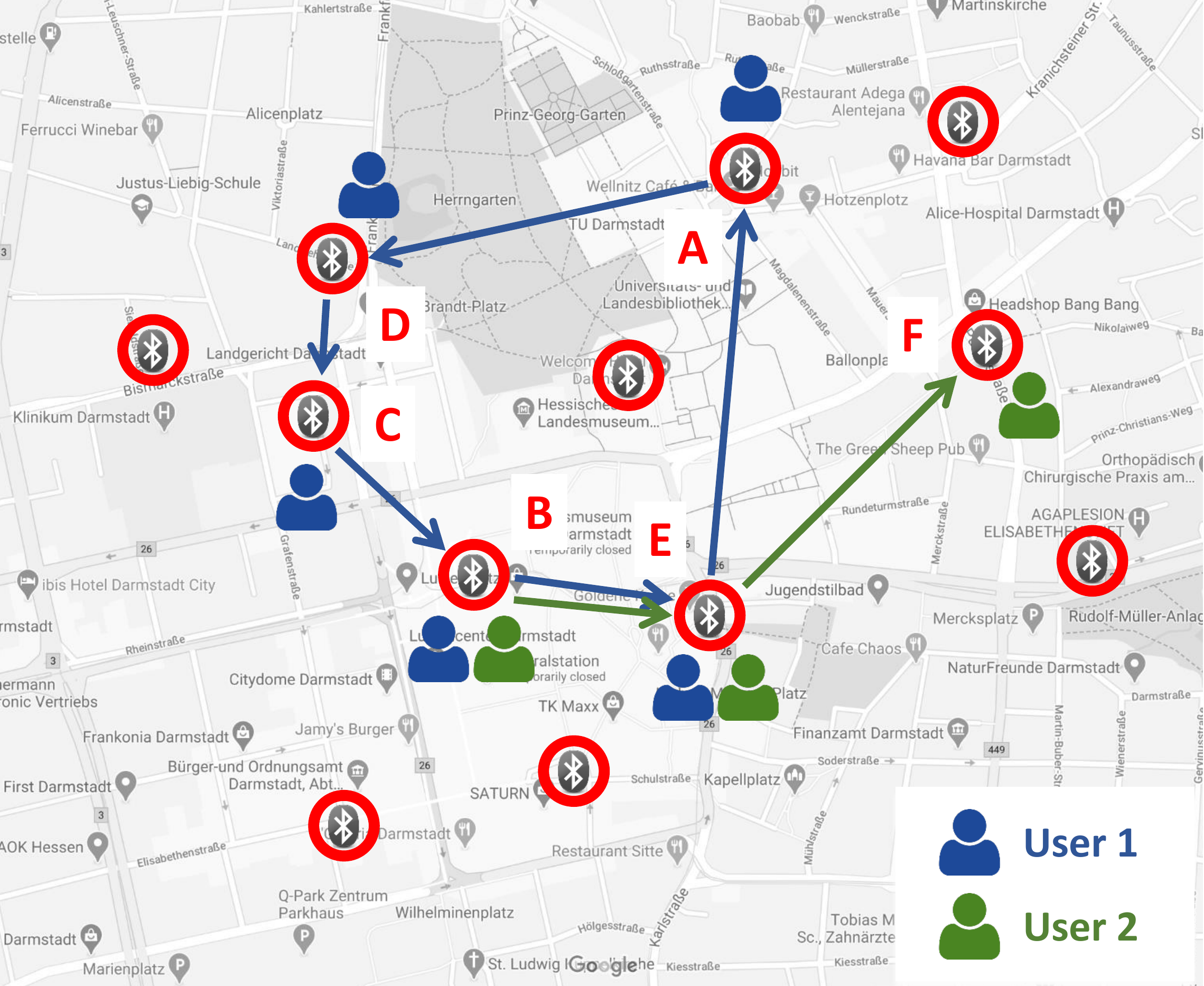}
    \caption{Movement profile of two infected users (User 1 in blue and User 2 in green) based on the observation points.}
    \label{fig:profiling-attack-route}
\end{figure}

Moreover, if we sort the locations that the users have visited in chronological order, we see that we can track the movements of each of the test users, as shown in Fig.~\ref{fig:profiling-attack-route}. Let (User $i$, $X$) denote the presence of User $i$ at location $X$. The sequence of observations of User 1 was as follows: (User 1, $A$), in a residential area, then (User 1, $D$) near a clinic and a pharmacy, (User 1, $C$) near the police station, (User 1, $B$) (Darmstadt city hall),  (User 1, $E$) (near to a pub) before concluding the round at the starting point with (User 1, $A$), corresponding again as mentioned, to a residential area. This may potentially indicate that the user may be living in this area. A similar tracing of locations is possible for User 2 who was first observed at (User 2, $B$), the city hall, after which the next observation  (User 2, $E$) happened near the pub, after which the final observation (User 2, $F$), was near a head shop and a sports gambling bookmaker.
With these observations about users and the associated timestamps, a significant amount of information can be gathered.
Since we know where users are, at which time, and how long they spend at each observed place, it is possible to aggregate relevant information from the users to potentially de-anonymize them.

For example, our experiment indicates that User 1 lives in the residential area near location $A$, and may have health and legal problems due to visiting the clinic and the police station. 
User 2 might be involved with the municipal administration and seems to like products available in a head shop or at a sports bookmaker.
Moreover, since  both users left location $B$ at about the same time, and even arrived at the pub (location $E$) at the same time, spent time there, and also left the pub at the same time, it is likely that these two users may have a social relationship.

We have conducted a series of experiments of different complexity. 
Our experiments demonstrate the power that the adversary gains by having access to RPI data of individual users. Since the TEKs change every 24 hours, traceability across longer time frames would initially not seem to be possible. However, since infected users upload TEKs of 14 days and since typical travel patterns of individual users show marked similarities even between different days (e.g., the typical commute pattern between home and workplace), it is possible to link and track at least some infected users for time periods significantly longer than the validity periods of individual TEKs (up to 14 days). This clearly will reveal even more personal information and activities of the targeted users and provide ample opportunities for using potentially available additional public information to de-anonymize the users in question. Moreover, de-anonymization becomes easier if the adversary has access to additional information about social relationships of users, e.g., the social graph of an online social network (OSN). This graph can be used to identify the infected users and their social contacts by comparing the social graph of the infected users obtained by the profiling attack to the OSN social graph~\cite{ji2015USENIXSecGraph, Radaelli2018SurveillanceNetworkedAge}.\footnote{Interestingly, a similar weakness was observed and criticized  \cite{dp3tAgainsPepppt} for the centralized tracing app, called PEPP-PT \cite{Peppptsp}, that enables the central server to build the social graph of infected individuals!} 

\textit{Experiments with the Corona-Warn-App:} We also conducted an experiment to confirm that the profiling attack is also applicable to the official German \emph{Corona-Warn-App} released on June 16, 2020. We captured RPIs from the \emph{Corona-Warn-App} without problems. We could also re-beacon these RPIs in our wormhole attack (cf. Section \ref{sec:relay-attack}).

\subsection{Case Study}
We consider a case in which an attacker seeks to identify persons infected with SARS-CoV-2 in Darmstadt, Germany. The attack would work best when useful side information is available, e.g., information about addresses of persons working in a particular place. For example, office addresses of the employees of the City of Darmstadt are available through the www.darmstadt.de website.

To capture coarse-grained movements of persons in Darmstadt, strategically-placed sensing stations need to be positioned in the city area. However, it is not necessary to place sensing stations over the full $122 \; km^2$ of the city area. This would require the ridiculous amount of $122 * 10 * 10 = 12,200$ sensing stations, even if one would place only one sensing station every $100$ meters in $1 \; km^2$. However, useful profiles can be produced even with less sensing stations.

\subsubsection{Persons Using Public Transport}
\begin{figure}[t]
\begin{subfigure}[c]{0.49\columnwidth}
    \centering
    \includegraphics[height=1.2\columnwidth]{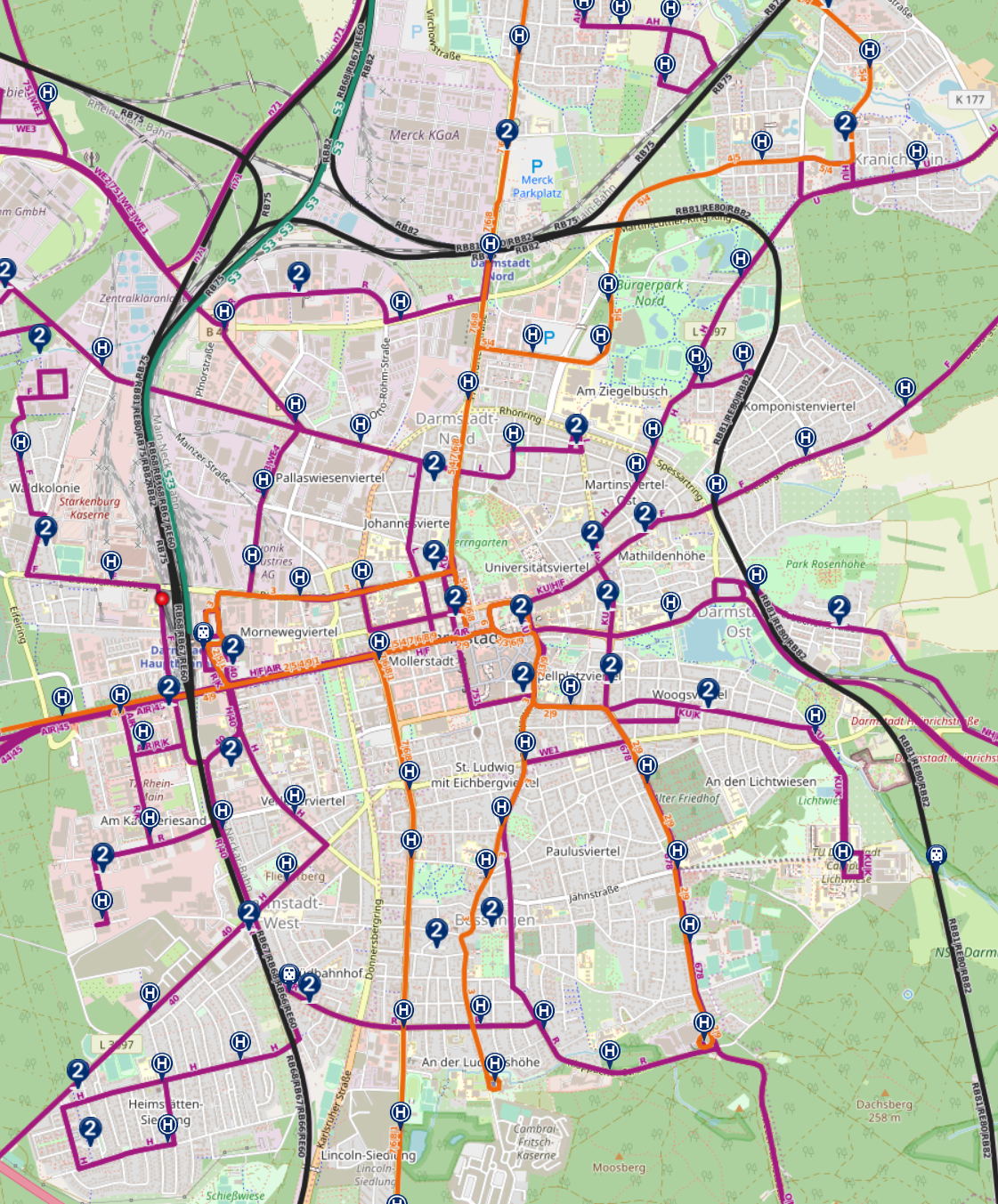}
    \subcaption{Train, tram, and bus lines}
    \label{fig:rmvtraffic}
\end{subfigure}
\begin{subfigure}[c]{0.49\columnwidth}
    \centering
    \includegraphics[height=1.2\columnwidth]{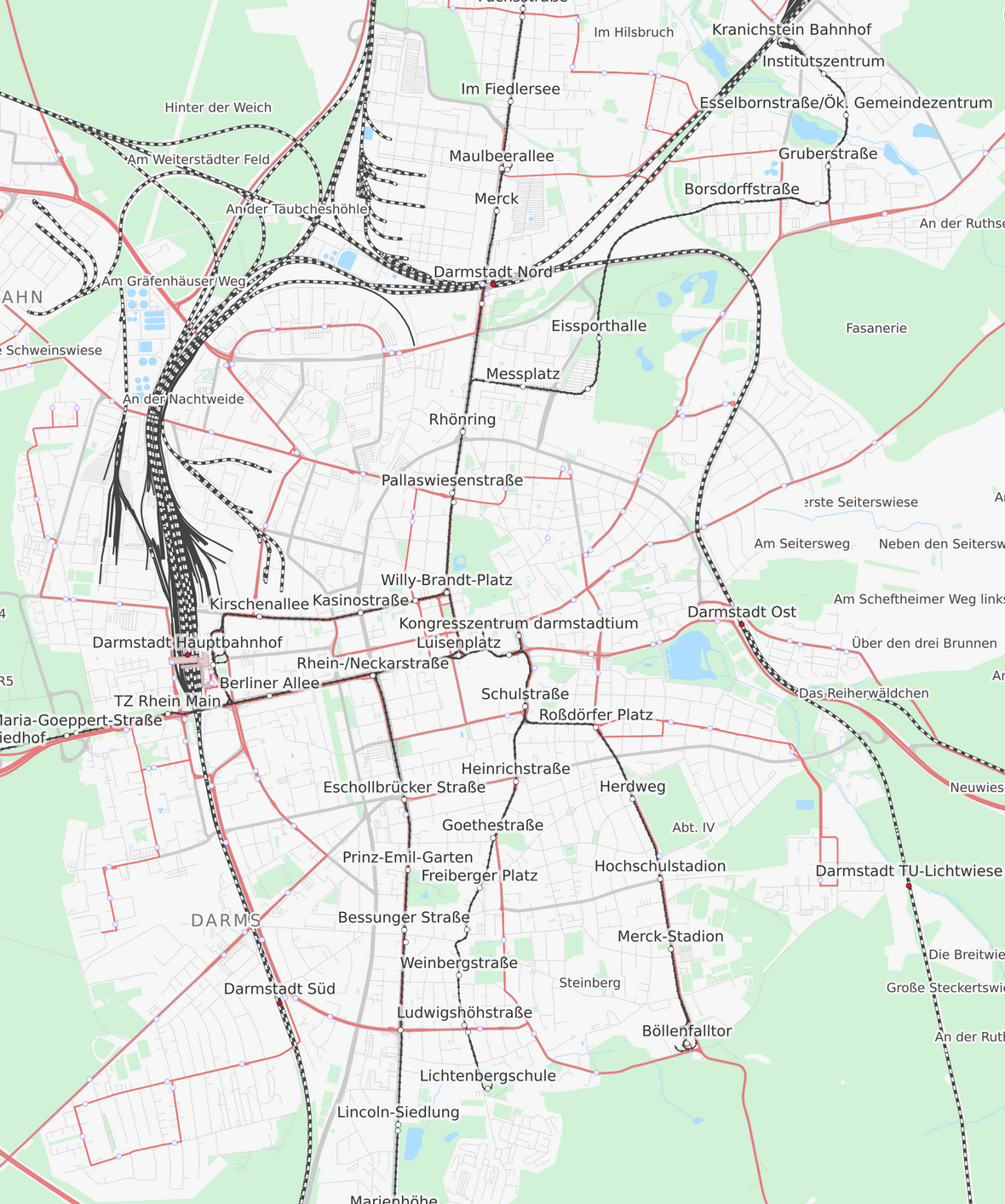}
    \subcaption{Main roads (cars \& railways)}
    \label{fig:cartraffic}
\end{subfigure}
\caption{Maps of Darmstadt showing main transport routes}
\label{fig:trafficmaps}
\end{figure}

There are $42 \; km$ of tram lines in Darmstadt servicing $75$ tram stops, depicted in Fig~\ref{fig:rmvtraffic}. Out of these, $54$ stops are in the city area of Darmstadt. The tram tracks run in a star shape around Luisenplatz on approximately six distinct axes. By placing $3-4$ sensing stations along these axes, one would be able to observe all tram passengers, from which direction they approach the Darmstadt city center or in which direction they leave. This would require around $25$ sensing stations.

There are approximately $150$ bus stops in the city area of Darmstadt. Bus lines move along approximately $20$ major bus routes. The bus routes are laid out such that they feed traffic to one of the major traffic centers. By placing $3-4$ sensing stations on each of these major bus routes, one could monitor the movements of the majority of passengers approaching these traffic centers. This would require $60-80$ strategically-placed sensing stations.

There are $9$ railway stops in Darmstadt servicing commuters arriving from farther away, shown as the blue railway icon in Fig.~\ref{fig:rmvtraffic}. 
Sensing stations needed to monitor movements of persons at railway stations will probably vary a lot, since the size of the stations is quite different; the main railway station is by far the largest. 
Sensing all persons entering and leaving the main railway station would require about $10-20$ sensing stations. Other train stations would require probably less, $2-5$ stations would be sufficient. Therefore, about $60$ sensing stations would be required to monitor persons entering or leaving Darmstadt by rail.

\subsubsection{Persons Using Private Transport}
There are five major and seven minor streets that persons typically use to enter Darmstadt by car. The streets are ordered in a star shape around the city center, as shown in Fig.~\ref{fig:cartraffic}. In addition, there are $12$ other streets inside the city for cross-connections between the major streets and individual city blocks. All of these streets are regulated using traffic lights forcing car drivers to stop several times while crossing the city area. By placing $4-5$ sensing stations (per direction) along the major streets, it would be possible to monitor persons entering Darmstadt by car. A similar number of sensing stations would be necessary to cover movements inside the city on the major streets. 
Thus, by using $200-250$ sensing stations, an adversary could have coarse-grained monitoring capability over all major street connections in Darmstadt.

There are a few places in Darmstadt that are heavily frequented and used by persons for changing connections of public transport. 
These places include Luisenplatz, Schloß, Hauptbahnhof, Nordbahnhof, Willy-Brandt-Platz, and Eberstadt Wartehalle; the busiest place is Luisenplatz. 
To have meaningful coverage over Luisenplatz, one would need to place $2-3$ sensing stations near all of the $6$ public transport stops on Luisenplatz, i.e., around 15 sensing stations. 
Similarly, monitoring the area around Hauptbahnhof would require placement of $1-2$ sensing stations at the 10 public transport stops in its vicinity, requiring around $20$ sensing stations in total. 
Other of the aforementioned connection points would likely require around 5 sensing stations per location. 
All in all, around $50$ sensing stations would be required.

\subsubsection{Summary}

\begin{table}[t!]
\caption{Estimate of how many sensing stations are needed to track persons using different kinds of transportation}
\centering
\renewcommand{\arraystretch}{1.2}
\begin{tabular}{lr}
\textbf{Location}    & \textbf{Sensing stations} \\ \hline
Trams       & 25              \\
Buses       & 60 - 80         \\
Railways    & 60              \\
Cars        & 200 - 250       \\
Pedestrians & 50              \\ \hline
\textbf{Total}       & 395 - 465 
\end{tabular}
\label{tab:trackingtransport}
\end{table}

To generate coarse-grained movement profiles of persons moving in and out of Darmstadt, an adversary would roughly require the sensing capabilities shown in Table~\ref{tab:trackingtransport}. 
This shows that by strategically placing a fairly modest ($<500$) number of sensing stations, an adversary would be able to monitor a large fraction of all traffic inside a major city of about $160,000$ inhabitants. 

The impact of the attack on a population level primarily depends on two factors: (i) the prevalence of the disease in the targeted area in the form of new daily infections, and (ii) the number of persons uploading their infection status via a contact tracing app. 
Consequently, in untargeted attacks, the adversary will focus the attack to areas where the daily infection rates are high to ensure maximal impact for the attack, assuming that many people in this area will upload their infection status via a contact tracing app. 
For targeted attacks, in which the adversary is targeting a particular organization like a business corporation, the population-level incidence plays a lesser role, since the target setting of the adversary is different. 
In targeted attacks, the adversary's focus is on obtaining detailed reconnaissance about any target persons in the targeted organization and therefore the overall number of persons affected by the attack plays a lesser role.

\section{Mind the Security GAP}
\label{wormhole}

In  this section, we present our 
real-world attack on the security of the GAP. 
The GAP design is vulnerable to relay-based wormhole attacks that can generate fake contacts resulting in a potentially large number of false alarms about contacts with infected persons, eventually leading to a high pressure on the public healthcare system administering COVID-19 tests. 
Relay attacks in the context of contact tracing apps have been discussed in a recent EU e-Health Network whitepaper \cite{eu2020mobile}, as well as in the context of PACT \cite{chan2020pact} and 
DP-3T \cite{vaudenay2020analysis,vaudenay2020centralized,pietrzak2020delayed}. Gvili \cite{gvili2020security} proposes countermeasures, such as time- and (geographical) cell-based verification of received messages or the need for bidirectional communication. Beskorovajnov et al. \cite{beskorovajnov2020contra} argue that every contact tracing protocol that does not incorporate a handshake mechanism or validation of time or location is vulnerable to relay attacks.

\subsection{Security Attack: Infection Alarms Going Viral}
\label{sec:relay-attack}

A wormhole attack is a particular type of relay attack.
Here, an attacker records messages at one physical location of a network, forwards them through a network tunnel to another  physical location, and re-transmits them there as if they had been sent at this location in the first place~\cite{hu2006wormhole}. 

\subsubsection{Goal and System Setup}
The goal of the attack is that two or more physical locations are combined into one large logical location.
If highly frequented physical locations, such as train stations, shopping malls etc., 
are logically linked to each other, 
an infected person can (falsely) be observed to have been in contact with other persons at remote locations. 
This can lead to a significant number of false alarms for people who might not have been in contact with anyone who is infected.
Furthermore, people might refuse to use the app, since the false positive rate seems to be too high, i.e., not helpful in contact tracing, but rather causing unnecessary work and confusion.

\begin{figure}
    \centering
    \includegraphics[width=1.0\columnwidth]{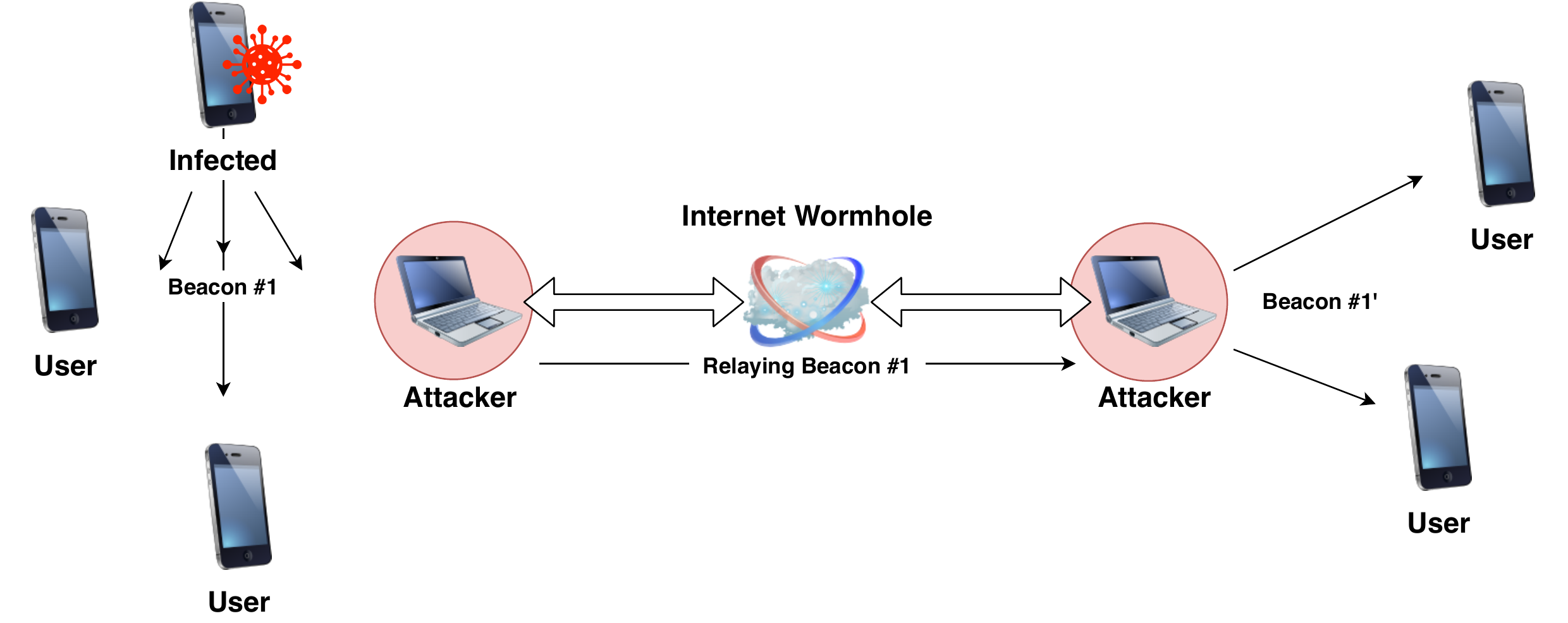}
    \caption{Wormhole attack setup to relay BLE beacons}
    \label{fig:covid_wormhole}
\end{figure}

Fig.~\ref{fig:covid_wormhole} shows the basic setting to perform a wormhole attack. 
The attacker uses (at least) two BLE enabled wormhole devices in different locations, each of them in physical proximity to mobile devices of potential victims, and records the BLE messages broadcast by the users' mobile devices at each location. 
The recorded BLE messages are then transferred, e.g., via an Internet link, in both directions between the wormhole devices of the attacker and are then re-broadcast to all mobile devices of users in the near vicinity of both locations of the wormhole devices.
Thus, these BLE messages seem to come from a local 1-hop neighbor.

GAP RPIs are only valid for a limited time interval.
An attacker 
performing a wormhole attack 
can add an expiration date to each received BLE beacon message before transferring it through the Internet link to other wormhole devices.
These wormhole devices can then re-broadcast the same BLE message over and over again until the expiration date is met.
While the used bandwidth of the Internet wormhole is kept to a minimum, the efficiency of the wormhole devices is maximized, since a single BLE message from a wormhole device can be resent multiple times to victims' devices for as long as the BLE beacon is valid.
While RPIs are valid 
by design 
for about 10 minutes, the specification grants a +/- two-hour tolerance time window between when RPIs should have been broadcast and when they are actually observed. This means that an attacker can replay each captured RPI for at least two hours in order to generate potential false positives.
Since BLE beacons are relatively small in size and the Internet provides fast communication, relaying is achieved in a matter of milliseconds.

\subsubsection{Experimental Results} 
The GAP API is only permitted to be used by developers or testers that are  authorized by governmental health institutions, since a special permission of either Google or Apple is required to use their GAP (\textit{Exposure Notification}) API. 
To validate that a wormhole attack was successful, we need to verify that the BLE beacons sent by a smartphone A to a wormhole device and transmitted over the network to another wormhole device are finally accepted by another smartphone B.

\paragraph{\emph{\textit{Experiment 1: DP-3T}}} 
Since the GAP approach is  heavily inspired by the DP-3T group and their approach, 
DP-3T can be considered as a substitute for GAP to show the success of a wormhole attack.
The original DP-3T implementation (later named \textit{prestandard} DP-3T\footnote{\url{https://github.com/DP-3T/dp3t-sdk-ios/releases/tag/prestandard}}) was available before the GAP API made it into Android and iOS.
In this \textit{prestandard} version, DP-3T generates TEKs as well as handles and stores RPIs as part of the app.
The operating system (iOS or Android) is in charge of handling the lower-level BLE communication, and the app provides the payloads required for BLE communication.

We built a multi-location wormhole for forwarding and rebroadcasting BLE messages based on off-the-shelf Raspberry Pis with integrated BLE and an Internet uplink, using our PIMOD tool \cite{hoechst2020pimod}.
These Raspberry Pis were connected using a central MQTT server as a back-end system for distributing the received BLE messages between each wormhole device. 
One of these Raspberry Pis functioned as a mobile node by using a battery pack and a mobile phone network uplink.
\begin{figure}
    \centering
    \includegraphics[width=1.0\columnwidth]{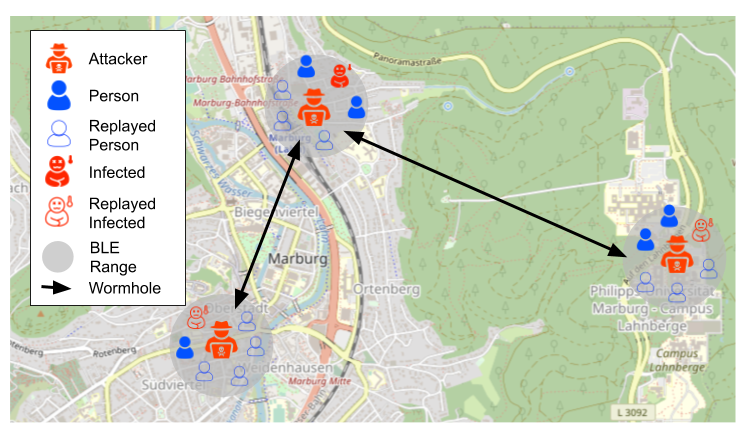}
    \caption{Wormhole attack in the city of Marburg}
    \label{fig:covid_wormhole_marburg}
\end{figure}
Our evaluation wormhole connects several physical locations in the cities of Marburg, Gießen, and Darmstadt, Germany. 
Part of this setup within the city of Marburg is shown in Fig.~\ref{fig:covid_wormhole_marburg}.
Each Raspberry Pi receives and records all BLE messages sent by surrounding users' iOS and Android smartphones and sends them to the MQTT wormhole server.
All other connected wormhole Raspberry Pis receive a copy of these BLE beacons from the MQTT wormhole server and rebroadcast them using their integrated BLE hardware. 
Each wormhole device is sender and receiver at the same time. 
Thus, this setup works in a bi- or multi-directional fashion.

In our tests, we used several iOS and Android smartphones with 
the corresponding implementation of 
the DP-3T prestandard \textit{SampleApp} at multiple physical locations in the three German cities mentioned above.
Our tests showed that the DP-3T prestandard \textit{SampleApp} is vulnerable to our wormhole attack. 
For example, we successfully established a logical contact between smartphones that were 40 kilometers apart in two cities without a real-world contact between the users of these smartphones. 
The logical contact between the smartphones was generated without any actions by the users required on their smartphones and without any physical interaction between the two individuals.

\begin{lstlisting}[
    style=log, 
    caption={Raspberry Pi with our wormhole implementation}, 
    label={lst:wormpi},
    float=h,
    captionpos=b,
]
Jun 09 20:45:13 wormpi-mr wormhole[472]: [provider    ] [INFO] [in ] [7E:09:47:A6:EE:7F] [Dp3t_ScanRequest] fd68
Jun 09 20:45:13 wormpi-mr wormhole[472]: [wormhole-out] [INFO] [7E:09:47:A6:EE:7F]       [Dp3t_ScanRequest] fd68
Jun 09 20:45:13 wormpi-mr wormhole[472]: [wormhole-in ] [INFO] [5A:A2:81:40:7A:B3]       [Dp3t_ScanResponse] fd68 6d:72:34:32:30:80:1d:62:d7:c9:ff:d0:71:a3:37:b0
Jun 09 20:45:13 wormpi-mr wormhole[472]: [provider    ] [INFO] [out] [5A:A2:81:40:7A:B3] [Dp3t_ScanResponse] fd68 6d:72:34:32:30:80:1d:62:d7:c9:ff:d0:71:a3:37:b0
\end{lstlisting}

In Listing~\ref{lst:wormpi}, an excerpt of the log of a running wormhole device, here called \textit{wormpi-mr}, is shown. 
The software on the wormhole device consists of a BLE controller and a beacon distribution task, called "provider".
The DP-3T prestandard \textit{SampleApp} used the BLE UUID \texttt{fd68}, which is correctly identified as \texttt{Dp3t\_ScanRequest} in the case of an empty payload and  \texttt{Dp3t\_ScanResponse} when the RPI is included.
In Lines 1-2, another wormhole device has submitted a \texttt{ScanRequest} beacon to the provider, indicated by "in", that is then broadcast by \textit{wormpi-mr's} BLE controller ("wormhole-out").
In Lines 3-4, a response is received by \textit{wormpi-mr's} BLE controller as "wormhole-in", which is then sent over the provider ("out") to all other wormhole devices.

\begin{figure}[htb]
\begin{subfigure}[c]{0.49\columnwidth}
    \centering
    \includegraphics[height=1.75\columnwidth]{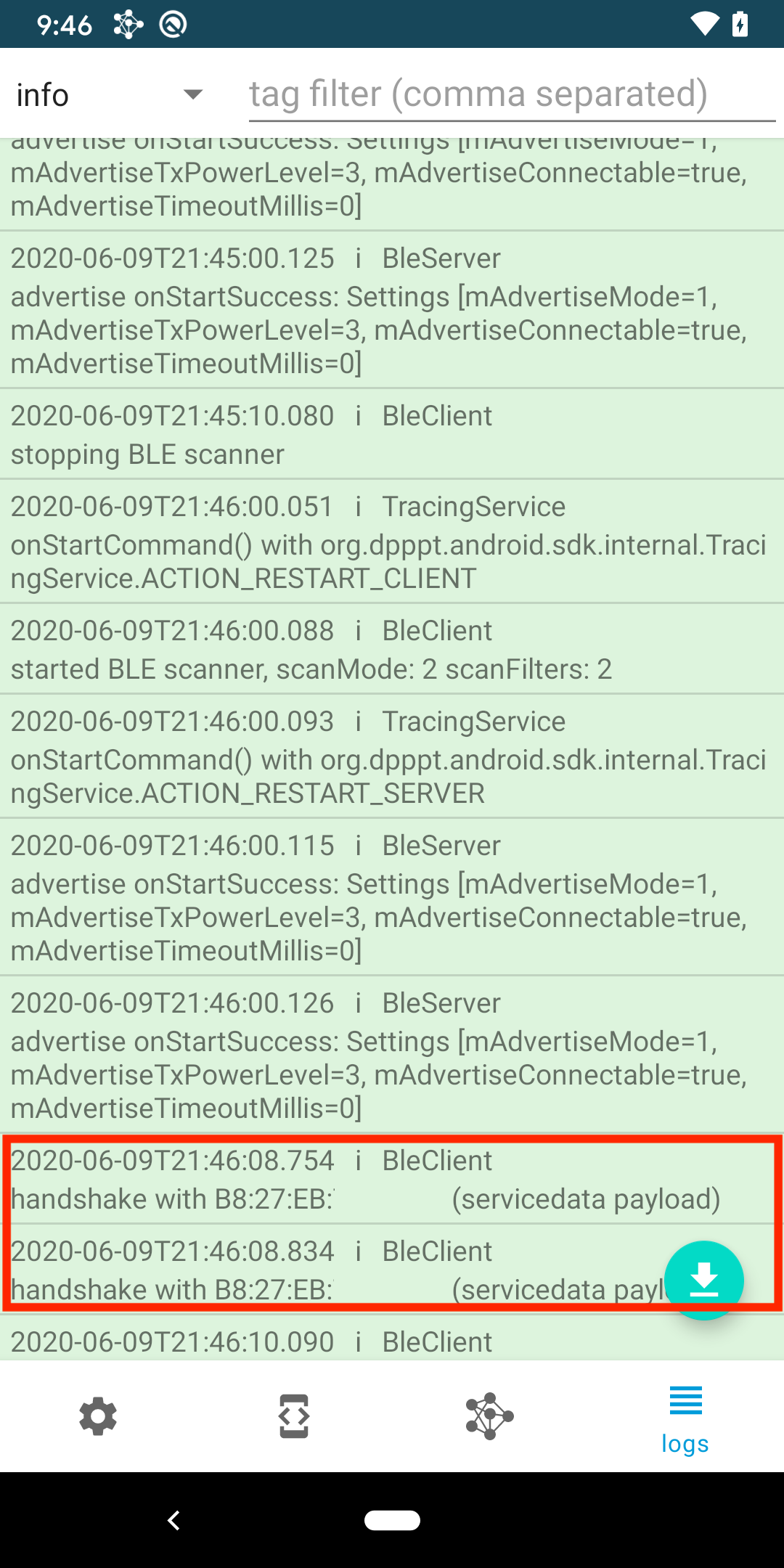}
    \subcaption{Android phone in Marburg}
    \label{fig:sampleapp-screenshots:android}
\end{subfigure}
\begin{subfigure}[c]{0.49\columnwidth}
    \centering
    \includegraphics[height=1.75\columnwidth]{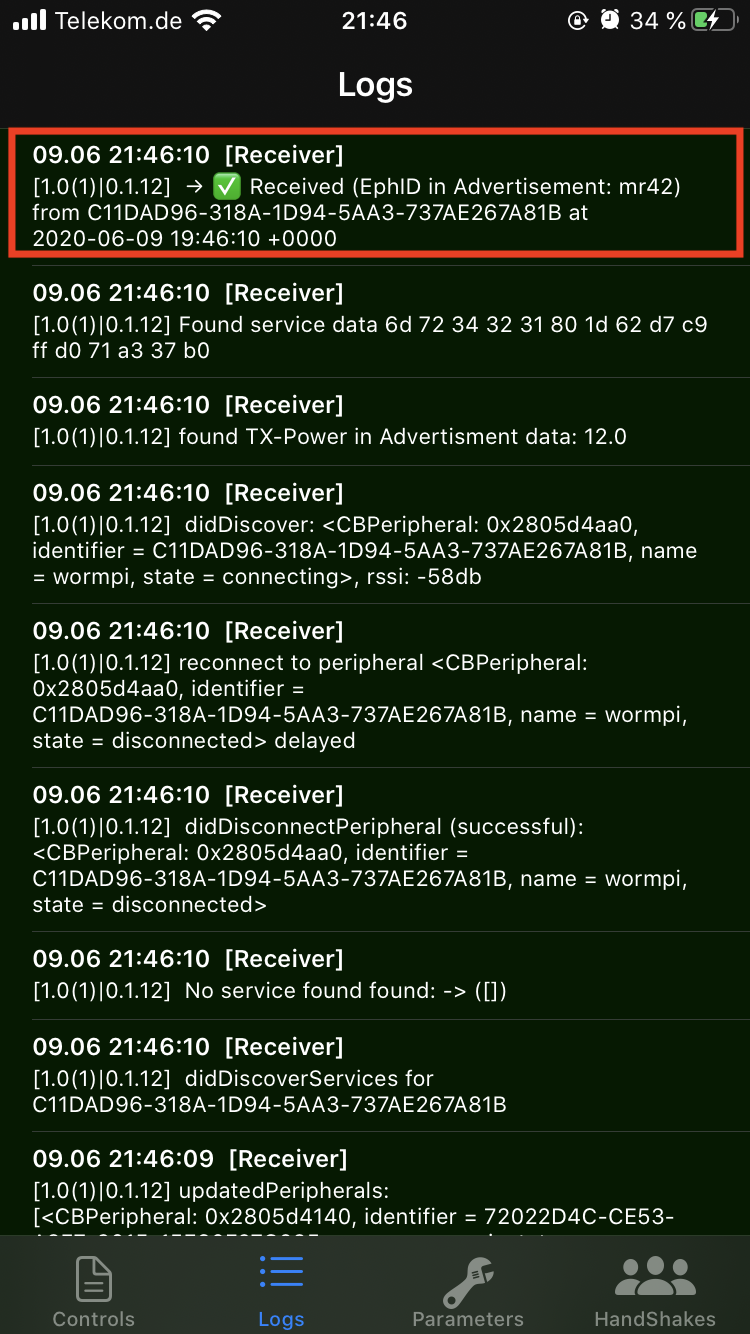}
    \subcaption{iOS phone in Gie\ss{}en}
    \label{fig:sampleapp-screenshots:ios}
\end{subfigure}
\caption{DP-3T prestandard \textit{SampleApp} instances with confirmed beacons transmitted through the wormhole "wormpi"}
\label{fig:sampleapp-screenshots}
\end{figure}

In Fig.~\ref{fig:sampleapp-screenshots}, two screenshots of running DP-3T prestandard \textit{SampleApp} instances on Android (Fig. \ref{fig:sampleapp-screenshots:android}) and iOS (Fig. \ref{fig:sampleapp-screenshots:ios}) are shown.
The experiments indicate that the execution of a wormhole attack was successful. 
The Android implementation on a smartphone located in Marburg (Fig.~\ref{fig:sampleapp-screenshots:android})
displays a handshake with the MAC address of the wormhole device (indicated by the rectangles in red), which in this experiment is the hardware MAC address of the used Raspberry Pi (abbreviated due to privacy reasons).
The iOS implementation on a smartphone located in Gie\ss{}en (Fig.~\ref{fig:sampleapp-screenshots:ios}) is less verbose, but also confirms receiving a beacon with the manually set ephemeral ID of "mr42" (i.e., the smartphone in Marburg; indicated by the rectangle in red), even though the smartphone is not in physical proximity of another smartphone running the DP-3T prestandard \textit{SampleApp}.

\begin{lstlisting}[
    style=log, 
    keywordstyle=\color{black},
    caption={Exposure notification confirming a received RPI}, 
    label={lst:logcatWrite},
    float=h,
    captionpos=b,
]
I/ExposureNotification: Scan device 6B:12:D2:1B:13:B5, type=1, id=31680EBB671454E1D7B03B2E96B98328, raw_rssi=-79, calibrated_rssi=-77, meta=919BAEA1, minutes_since_last_scan=1594815319 [CONTEXT service_id=236 ]
I/ExposureNotification: BleDatabaseWriter.writeBleSighting, id=31680EBB671454E1D7B03B2E96B98328 [CONTEXT service_id=236 ]
\end{lstlisting}

\paragraph{\emph{\textit{Experiment 2: German Corona-Warn-App}}}
We also validated our results using the Android version of the official German \textit{Corona-Warn-App} released on June 16, 2020.
As shown in Listing \ref{lst:logcatWrite}, the GAP of the \textit{Corona-Warn-App} stores RPIs transmitted using the wormhole. 
Since we could not get permission to access the GAP API,
we used a TEK from the official server of the \emph{Corona-Warn-App}, derived multiple RPIs and injected these into the wormhole. Since the derived RPIs do not contain the \textit{Associated Encrypted Metadata} (AEM) that would normally be broadcast with an RPI, we had to derive the AEM, too. 

To validate that our RPI derivation is correct, 
we used Frida\footnote{\url{https://frida.re/}} on a rooted Google Pixel 3 smartphone to extract all TEKs stored on this device. Using a known TEK together with numerous RPIs and their corresponding time slots, we could validate that our RPI derivation works correctly. Additionally, this approach allowed us to decrypt the AEM for the RPIs. This non-encrypted metadata was then used to generate valid AEM for the derived RPI. 
These RPIs are only valid for the time of the initial creation, therefore we had to change the system time of the receiving device. Otherwise, the device would not be able to match the keys against the uploaded TEK. Changing the system time is only necessary for our validation purposes.

\begin{lstlisting}[
    style=log, 
    keywordstyle=\color{black},
    caption={Automated generation of valid RPIs}, 
    label={lst:rpigeneration},
    float=h,
    captionpos=b,
]
D/BackendManager: [ TEK: fd3df1b125a21a28f1d7746fd5a46538 ] encrypted Metadata for 9386bead6a0212d6205c665db64ccfe4 = a4e4489c @ Time/Date(Tue Jul 07 00:00:00 GMT+02:00 2020 | 2656788) - Full BLE-Payload: 93:86:be:ad:6a:02:12:d6:20:5c:66:5d:b6:4c:cf:e4:a4:e4:48:9c
D/BackendManager: [ TEK: fd3df1b125a21a28f1d7746fd5a46538 ] encrypted Metadata for 3b65333a5383d8c4d6344672a14963de = 3d167031 @ Time/Date(Tue Jul 07 00:10:00 GMT+02:00 2020 | 2656789) - Full BLE-Payload: 3b:65:33:3a:53:83:d8:c4:d6:34:46:72:a1:49:63:de:3d:16:70:31
\end{lstlisting}

The payload for an exposure notification (e.g., see Listing \ref{lst:rpigeneration}) was then injected into the wormhole. On the receiving side, the official \textit{Corona-Warn-App} was installed on a device, and its system time was set to the corresponding interval of the RPI. During the approximately 15 minutes this experiment took, the GAP API received and stored exposures several times (similar to Listing \ref{lst:logcatWrite}). 
Afterwards, the receiving device was set back to the correct time and communication with the server containing the TEKs of known infected app users was re-enabled.
Since the \textit{Corona-Warn-App} does not submit the TEK for the same day to the GAP API twice and the list of keys is signed, we had to ensure that the app could not receive the TEKs for the specific date before the experiment finished.
Since the app checks for an existing Internet connection, 
we used a proxy to block requests to the server during our test. In this way, the test for an existing Internet connection succeeded, but the app could not retrieve TEKs.

\begin{figure}[htb]
\centering
\includegraphics[height=0.98\columnwidth]{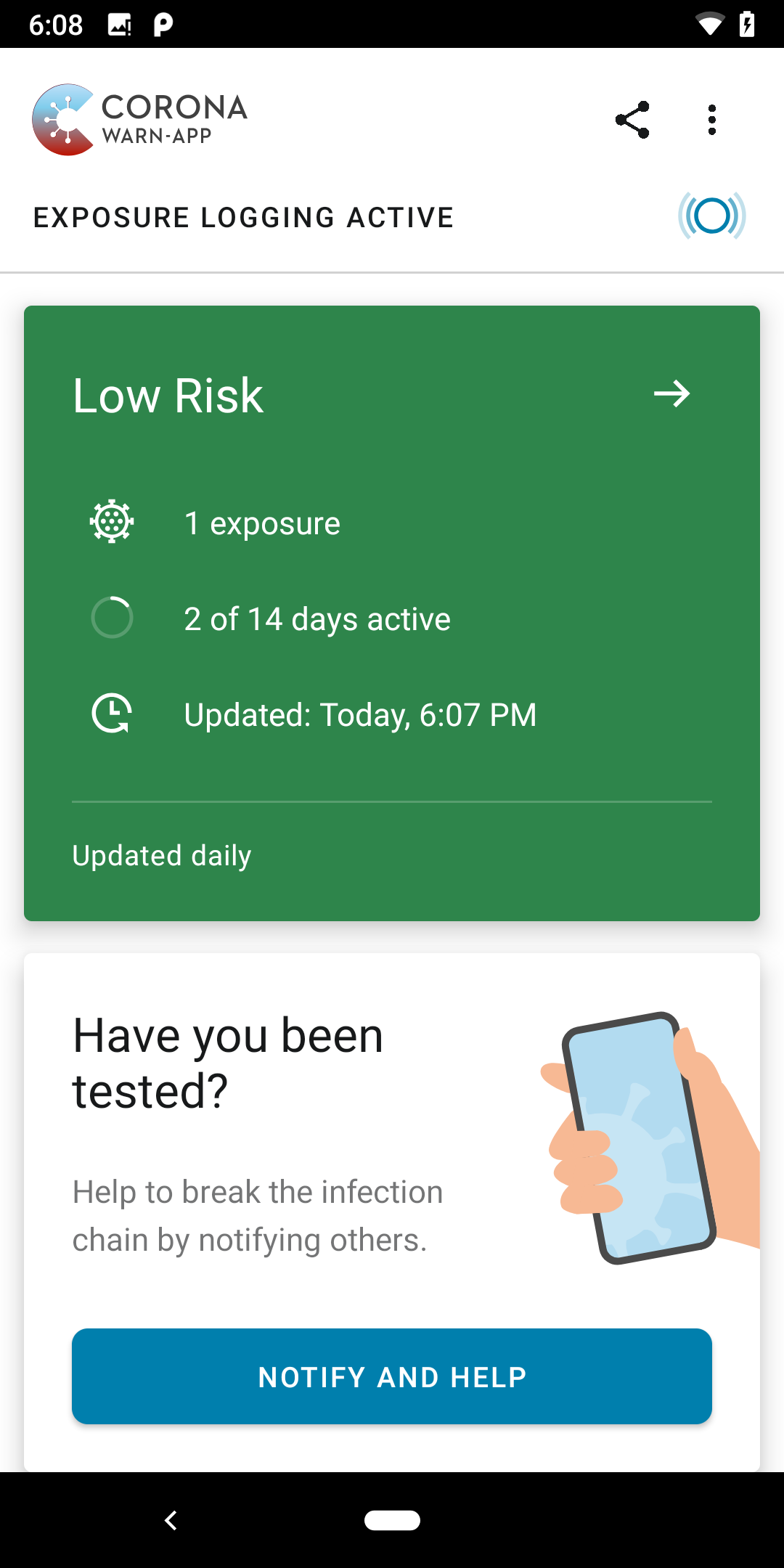}
\caption{Official German \textit{Corona-Warn-App} with a positive exposure transmitted through the wormhole}
\label{fig:cwa-screenshots}
\end{figure}

As shown in Figure \ref{fig:cwa-screenshots}, the \textit{Corona-Warn-App} reports a single exposure. The low risk level shown is due to the low \textit{Transmission Risk Level} of the chosen TEK and the metadata of the transmitted RPI. Although the transmitted RPI should not have been valid for regular devices at the time of the broadcast, these values were chosen on purpose to reduce the impact for people who might have been present in the surrounding area. 

It should be noted that the described steps are only necessary to validate that the \textit{Corona-Warn-App} is indeed vulnerable to our wormhole attack. The attack itself does not require any modification of software running on the device.

\subsection{Technical Limitations}
\label{subsecteclimit}
The GAP distributes beacons using the newer BLE standard~\cite{BluetoothCore52}, 
allowing physical transmission speeds of up to 1 Mbps.
To allow more robust transmissions, the physical layer also offers representations with 2 and 8 symbols, resulting in transmission speeds of 500 kbps and 125 kbps, respectively, which are not discussed here for the sake of simplicity.

The payload of the GAP Exposure Notification service has a combined size of 26 bytes \cite{AppleExposureBluetoothSpecification}.
A GAP beacon with a size of 26 bytes is sent via an undirected advertising event, resulting in an advertisement size of 39 bytes and a packet data unit size of 47 bytes~\cite{BluetoothCore52}.
With 1 Mbps, a single advertisement with a size of 47 bytes (= 376 bits) results in an on-air time of $376 \mu{}s$. 
In addition, an inter-frame space of $150 \mu{}s$ is required after each advertisement. 
Hence, a theoretical maximum rate of $10^6 \mu{}s / \frac{(376 \mu{}s + 150 \mu{}s)}{packet} = 1,901 \; packets/s$ can be sent using BLE 4.0 advertisements according to the GAP specification. 

In a real world setting, there are several factors that significantly reduce the theoretical maximum rate, such as:
\begin{itemize}
\item BLE advertisements are sent using three BLE channels; receivers need to hop between these channels;
\item connection intervals forced by the device vendor;
\item distance between receiver and sender; BLE has a transmission power of 10 mW (i.e., a distance of up to 40 meters), and Class 1 BLE devices have up to 100 mW (i.e., a distance of up to 100 meters);
\item interferences and collisions.
\end{itemize}

To evaluate the impact of these factors, we set up a test environment consisting of:  HackRF (sender, repeater), Raspberry Pi (receiver), Eve PowerPlug (BLE interference, distance about 2 meters), Ubiquity AP nanoHD (WiFi interference on 2.4 Ghz, distance less than 2 meters, 100\% load and transmission power). In this test environment, we received only about 4.3\% of the theoretical maximum of BLE advertisements per second (i.e., 82 BLE advertisements/s) using a consumer grade BLE receiver with factory default settings while sending on a single BLE advertisement channel. We also discovered that most of our tested devices submit BLE advertisement packets once every two seconds. Furthermore, most BLE devices only accept packets for a short period of time every few minutes.

In an indoor test with an active interfering WiFi hotspot and several interfering BLE devices, the maximum distance was reduced to below 10 meters with direct line of sight between sender and receiver. However, using a signal repeater, we were able to increase the distance up to 50 meters.

\subsection{Attack Scenario: Opportunistic Linking}
\label{subsecoplink}
To increase the probability of a successful wormhole attack, an attacker can
\begin{enumerate} 
\item  increase the number of collected BLE advertisements by selecting a highly frequented area with a high acceptance rate of the particular contact tracing app, and by increasing the number of deployed wormhole devices;
\item  increase the probability that one of the relayed RPIs belongs to a person who will be tested positive 
for COVID-19 
and uploads his or her TEKs by selecting an area with a high probability of infected persons.
\end{enumerate}

\subsubsection{Selecting a highly frequented area for RPI collection} We conducted experiments at the central train station in Frankfurt (Main), Germany, on the 1st of August 2020.
We moved around the train station with a OnePlus 7 Pro smartphone while changing trains, and waited in the main hall while collecting GAP BLE advertisements using the RaMBLE\footnote{\url{https://play.google.com/store/apps/details?id=com.contextis.android.BLEScanner}} app.
Since the people moved around at the distinct locations where we collected GAP BLE advertisements, we argue that the number of unique RPIs is roughly identical to the number of distinct users.  
During the first run, 549 unique RPIs were collected in 00:25:49 h, i.e., 21.26 collected BLE advertisements per minute. 
During the second run, 142 unique RPIs were collected in 00:04:40 h, i.e., 30.43 collected BLE advertisements per minute. 

In less populated environments, (a) inside an isolated examination room of the pulmology ward of the University Hospital of Heidelberg (only open to emergency patients and medical staff)
and (b) while driving in a car on the German highway A5 from Gie\ss{}en to Mannheim, we collected 95 (300) unique RPIs in 01:32:00 h (01:51:00 h) (i.e., 1.02 (2.70) BLE advertisements per minute).

To estimate the impact of our wormhole attack, we assume that there are 
(i) 5.1 reported infections among 100,000 persons per week (i.e., the average value for Germany published on August 1, 2020) and 
(ii) 30.43 collected unique BLE advertisements from distinct smartphones per minute by each wormhole receiver (according to our tests in the Frankfurt central train station).

First, we address the question of how many RPIs would be required to receive an average of one positive RPI.
Since infected users upload their TEKs of the last 14 days, the doubled weekly incidence value is a suitable estimator. 
Hence, with assumption (i) one out of $1 / (5.1 / 100,000 / 7 * 14) \approx 9,804$ received RPIs will be positive.
The average validity period of a received RPI can be estimated by halving the general validity period of 10 minutes, since some RPIs will be  received just after creation, while others will have almost expired.
To get access to one valid RPI at any given point in time, $9,804$ RPIs $/$ $30.43$ RPIs per minute per device $/$ $5$ minutes $\approx¢$ $65$ wormhole devices would be required.

By changing assumption (i) to 45.4 reported infections among 100,000 persons per week (i.e., the average value for Germany published for week 42 of 2020), the corresponding number of required wormhole devices at distinct locations drops down to about $1 / (45.4 / 100,000 / 7 * 14) / 30.43 / 5  \approx 8$.

The relatively low numbers of wormhole devices required for an attacker suggests that the attack can be carried out without much effort.
However, in the case of the German \emph{Corona-Warn-App}, the calculated risk score is set to zero, if the encounter period with an infected person is shorter than 10 minutes.
Hence, a successful attack would require an attacker to observe potentially infected people for a period of at least 10 minutes.

\subsubsection{Selecting an area with a high probability of infected persons}

We assume that a location with a high probability of infected persons can be used by the attacker, e.g., the COVID-19 Testing Center near Frankfurt (Main), Germany, performing a current maximum of 300 tests per hour.
Let us further assume that 3.62\% of the tests are positive (i.e., valid for Germany in week 42 of 2020),
and 9.84\% of the infected persons share their infection status (based on the calculations of submitted TEKs in relation to the overall reported infections in Germany (6.473 of 65.410 in week 41 and 42 of 2020)). 

Based on these assumptions, an attacker would  be able to observe an average of 10.86 infected people per hour, 
of which 1.07 upload their infection status through the app.
To generate a high risk warning, an attacker would select a test center where an individual can be observed for more than 10 minutes.

In the middle of October 2020, Germany was a relatively low risk country, but in other countries with a higher test-positive rate, these calculations look differently. 
For example, in Mexico in the middle of October 2020, 41.0\% of the tests were positive\footnote{\url{https://ourworldindata.org/coronavirus/country/mexico?country=~MEX} October 14, 2020)}. 
Using this rate (hypothetically) in the calculation for the test center in Germany, 123 infected persons could be observed per hour, of which 12.10 persons would upload their infection status, i.e.,
a positive RPI roughly every 5 minutes is obtained.

Since each RPI remains valid for a period of 120 minutes, an attacker can repeat RPIs even when the infected person is not within reach of the wormhole anymore.
If an attacked smartphone receives these RPIs over 10 minutes, the GAP would register 
$1.07 / 60 * 120 = 2.14$ infected persons in Germany and $12.10 / 60 * 120 = 24.20$
infected persons in Mexico in close proximity and would probably trigger high risk warnings.
Apparently, the impact of the attack can be limited by shrinking the 2-hour validity period of RPIs. 

\subsection{Attack Scenario: Targeted Attack}

In this scenario, an attacker has the possibility to submit his or her own TEKs at will. Depending on the local implementation of the GAP, this can require the acquisition of a valid TAN for uploading TEKs to the central governmental back-end server. The wormhole will solely be used as a publishing device. 

We will focus on the German \textit{Corona-Warn-App} as an example, for which a TAN can be obtained in different ways. 
Our team had contact with a supplier of TANs in a dark web underground community, therefore we assume that there is a "market" for TAN keys in exchange for money.
It is also possible to request a valid TAN by uploading a (forged) diagnostic report directly in the \textit{Corona-Warn-App} itself, which will be issued after a manual check by the hotline phone support team of the app.

Based on our collected real world data (see Section \ref{subsecoplink}), we can broadcast a positive BLE advertisement to about 30.43 mobile devices per minute $\times$ 60 minute $\approx$ 825 mobile user devices per hour per wormhole device. 

Currently, the last 14 days of exposure are considered for a warning by the GAP. 
In this case, a single wormhole device would be able to submit $1,825 * 14 * 12 = 306,600$ registered, positive RPIs to other mobile user devices during daytime (12 hours) for 14 days.
In the case of the German \emph{Corona-Warn-App}, RPIs over a period of 10 minutes are required to trigger a high risk warning. 
Thus, only the subset of people successfully attacked for more than 10 minutes and within a low distance and during the days of high infectiousness will get a high risk warning.

\section{Conclusion}
\label{conclusion}

We demonstrated that in real-world scenarios the current GAP design is vulnerable to (i) profiling and possibly de-anonymizing infected persons, and (ii) relay-based wormhole attacks to generate fake contacts that may affect the accuracy of GAP-based contact tracing apps. 

Since the fundamental vulnerabilities are based on the GAP itself, all GAP-based apps 
face the same issues. In fact, we demonstrated that the official German \emph{Corona-Warn-App} 
in its current version 
is vulnerable to our wormhole attack. 

In the future, a revision of the GAP might be required 
to implement countermeasures against our attacks, such as reducing the 2 hour RPI and 24 hour TEK validity period \cite{AppleExposureBluetoothSpecification}. 
However, more sophisticated countermeasures, such as handshake mechanisms or validation of time and/or location \cite{beskorovajnov2020contra,gvili2020security},
would probably pose additional threats to privacy and security of GAP-based apps and could lead to increased resource demands, particularly in terms of battery consumption.

\section*{Acknowledgment}
This research work has been funded by the Deutsche Forschungsgemeinschaft (DFG) – SFB 1119 – 236615297.

\bibliography{IEEEabrv,literature}{}
\bibliographystyle{IEEEtran}

\end{document}